\documentclass[aps,preprint,showpacs,superscriptaddress]{revtex4}
\usepackage{graphicx}
\usepackage{amsmath,amssymb,latexsym}
\bibstyle{apsrev.bib}

\def\WIDTHC{\textwidth}  
\newcommand{\beq}{\begin{equation}}
\newcommand{\eeq}{\end{equation}}

\begin{document}

\title{On the size of knots in ring polymers.}

\author{B. Marcone}
\affiliation{Dipartimento di Fisica, Universit\`a di Padova,
I-35131 Padova, Italy.}
\author{E. Orlandini}
\affiliation{Dipartimento di Fisica and Sezione CNR-INFM, 
Universit\`a di Padova, I-35131 Padova, Italy.} 
\affiliation{Sezione INFN, Universit\`a di
Padova, I-35131 Padova, Italy.}
\author{A. L. Stella}
\affiliation{Dipartimento di Fisica and Sezione CNR-INFM, 
Universit\`a di Padova, I-35131 Padova, Italy.} 
\affiliation{Sezione INFN, Universit\`a di
Padova, I-35131 Padova, Italy.} 
\author{F. Zonta}
\affiliation{Dipartimento di Fisica, Universit\`a di Padova,
I-35131 Padova, Italy.}

\begin{abstract}

We give two different, statistically consistent definitions of the
length $\ell$ of a prime knot tied into a polymer ring. In the
good solvent regime the polymer is modelled by a self avoiding
polygon of $N$ steps on cubic lattice and $\ell$ is the number of
steps over which the knot ``spreads'' in a given configuration. An
analysis of extensive Monte Carlo data in equilibrium shows that
the probability distribution of $\ell$ as a function of $N$ obeys
a scaling of the form $p(\ell,N) \sim \ell^{-c} f(\ell/N^D)$, with
$c \simeq 1.25$ and $D \simeq 1$. Both $D$ and $c$ could be
independent of knot type. As a consequence, the knot is weakly
localized, i.e. $\langle \ell \rangle \sim N^t$, with $t=2-c
\simeq 0.75$. For a ring with fixed knot type, weak localization
implies the existence of a peculiar characteristic length $\ell ^
\nu \sim N^{t \nu}$. In the scaling $\sim N^{\nu}$ ($\nu \simeq
0.58$) of the radius of gyration of the whole ring, this length
determines a leading power law correction which is much stronger
than that found in the case of unrestricted topology. The
existence of such correction is confirmed by an analysis of
extensive Monte Carlo data for the radius of gyration. The
collapsed regime is studied by introducing in the model
sufficiently strong attractive interactions for nearest neighbor
sites visited by the self-avoiding polygon. In this regime knot
length determinations can be based on the entropic competition
between two knotted loops separated by a slip link. These
measurements enable us to conclude that each knot is delocalized
($t\simeq 1$).
\end{abstract}


\maketitle

\section{INTRODUCTION}

Various forms of topological entanglement play a fundamental role
in determining equilibrium and dynamical properties of single chain
and multi-chain polymeric systems~\cite{DeGennes,Doi}, with
relevant consequences also for biological matter. For instance,
the presence of a knot can be an obstacle to the processes of
duplication and segregation of DNA in bacterials~\cite{Alberts}.
Indeed, there exist topoisomerase enzymes whose function is
precisely that of controlling the topology of circular
DNA~\cite{Rybenkov,Shaw}. The knots and links which are ubiquitous
in higher molecular multi-chain melts and solutions can profoundly
affect properties of such systems like viscosity or resistance to
rupture \cite{Pieranski}. Knots can even be found in the native
state of some proteins
\cite{Taylor,Lua,Virnau2006,Marconeprotein}. and may play an
important role in their stabilization with respect to denaturating
agents and in their folding dynamics.

\begin{figure}[tbp]
\includegraphics[angle=0,width=\WIDTHC]{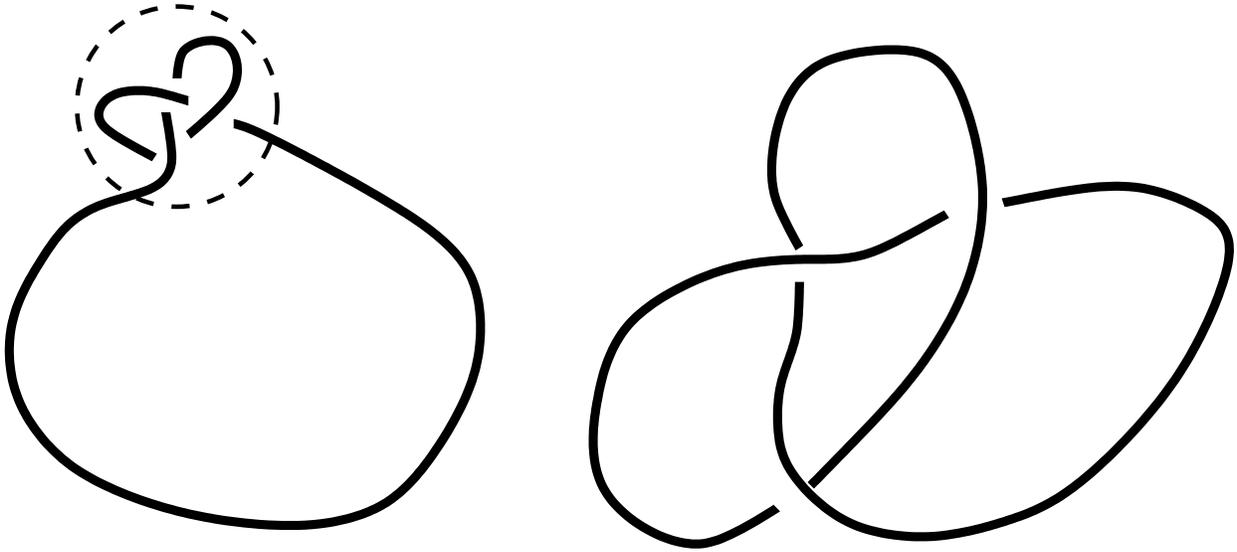}
\caption{On the left we see a tight knot: it is easy to say that the knotted
arc is that within the small sphere (dashed circle). 
On the right the knot
is delocalized within the curve.}
\label{figure1}
\end{figure}

The description of the consequences of topological entanglement
in polymer physics poses theoretical and numerical challenges
which only relatively recently started to be faced with some
success \cite{SumWhi} \cite{Pippenger}.
An interesting issue addressed in the last years is
that of establishing whether knots tend to be ``spread'' over the whole polymer
or ``localized'' within a short portion of the chain (Fig.~\ref{figure1}).
If properly quantified, the degree of localization of (prime)
knots is expected to play an important role in the discussion of both
equilibrium and dynamical properties of knotted
macromolecules. For example, if the knot is localized to some degree
in a long ring, the logarithmic correction to the ring entropy
per monomer should drastically change with respect to that
of the unknotted case \cite{OTJW98}.
On the other hand, the knot could behave in such a way that
its average ``length'' grows with the $t$-th power $(0 <t< 1)$ of
the total ring length $N$. Corrections to scaling associated to
this length should then be expected for the long chain behavior
of measurable quantities like the gyration radius \cite{Marcone}.
These corrections should be detectable as peculiar of rings with
prime knot, but could not be predicted within the framework
of approaches like the field theoretical renormalization group,
which treats only the case of ``phantom" ring polymers with 
unrestricted topology.
The size of the knot in a
DNA ring should also strongly affect the action of topoisomerases or the
mobility of the ring in gel electrophoresis experiments \cite{Stasiak}.
Furthermore, recent experiments of
DNA micromanipulation by optical tweezers have shown that it is
possible to tie specific knots into the macromolecule \cite{Aray}
and to observe their motion within a viscous
solution~\cite{QuakePRL}. For this problem the knowledge of the length
of the entangled region is essential, since it directly affects
the knot diffusion coefficient \cite{QuakePRL}.

\begin{figure}[tbp]
\includegraphics[angle=0,width=\WIDTHC]{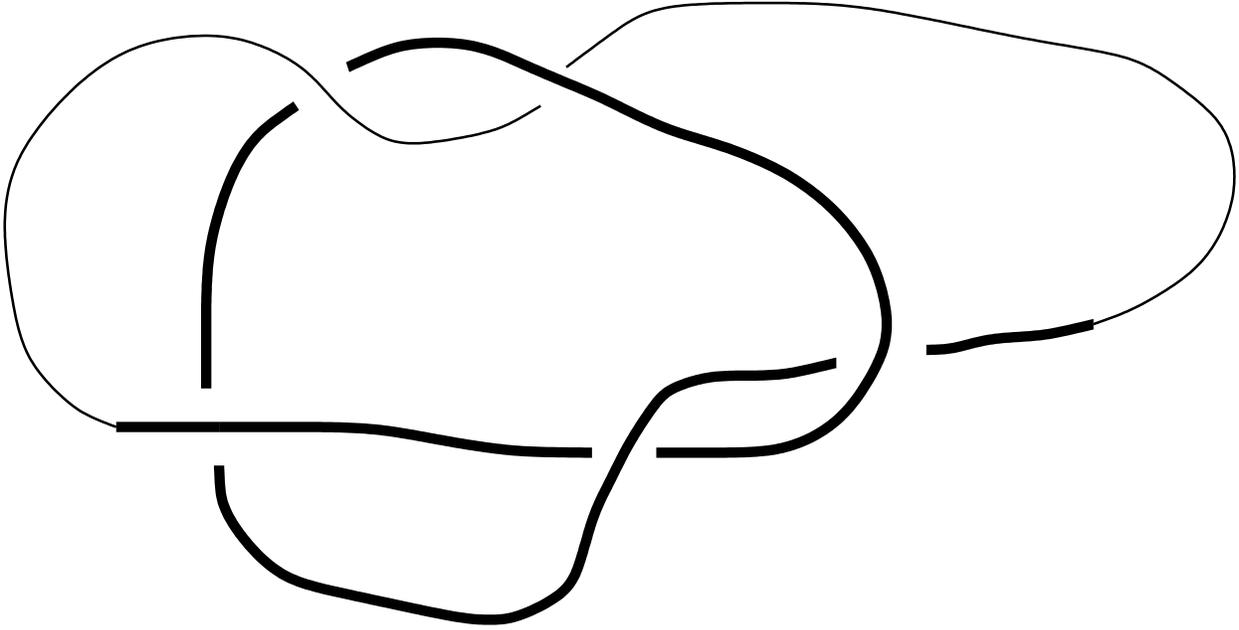}
\caption{The thicker arc, once extracted from the rest
of the curve, seems to be knotted while the whole curve is not.}
\label{tris}
\label{figure2}
\end{figure}

In spite of some early indications that prime knots in ring polymers in
good solvent are likely to be localized in small portions
of the chain \cite{OTJW98,Katrich}, sufficiently direct and
quantitative evidence of this property remained for long a major challenge.
This is mainly due to the difficulty of locating the knot of a closed
curve in a consistent way.
A possible procedure is that of isolating a trial open
portion of the curve and of checking whether the new ring
obtained by joining its extremes with a topologically
``neutral'' closure still contains the original knot, or not. The knot length
should then be identified with that of the smallest portion for which the knot
remains. Such procedure relies on the notion of knotted arc that is not
well defined mathematically \cite{Janse92}. Indeed, since knots are
embeddings of \emph{circles} \cite{Rolfsen}, in a strict mathematical sense
no open string can be knotted: continous transformations acting on such
string can always bring it into an untangled shape. For a general
closed curve knottedness is a global property: we can not state that
a portion is knotted, but only that the whole curve is
(Fig.~\ref{figure2}). Nevertheless, as we show here,
when dealing with a whole
sample of closed curve configurations the notion of
knot length may acquire a physical meaning, at least in a statistical sense.


A definition of knot length may be much more easy to give for
flat knots \cite{Guitter}, i.e. knots in ring polymers that are confined
in two dimensions \cite{Metzler02}. Physical examples include
polymer rings adsorbed on a plane by adhesive forces \cite{Maier}
or macroscopic necklages flattened under gravity onto
a vibrating plane \cite{Ben-Naim}.
The configurations of such adsorbed rings would be
similar to the planar projections used in knot theory to compute
topological invariants and classify knots~\cite{Rolfsen}. Under the
simplifying assumption that the number of overlaps
is restricted to the
minimum compatible with its topology (for example 3 for the
trefoil knot), the length of the hosted flat knot can be unambiguously
defined and its statistical behavior, as a function of the number of
monomers of the ring, can be studied analytically~\cite{Metzler02}  and
numerically \cite{OSVPRE}. In the good solvent regime flat knots were found
to be strongly localized.
This approach has also been
extended to polymer rings that undergo collapse from the
swollen (high temperature) to the compact (low temperature) regime
and it was found that globular flat knots are delocalized $(t \simeq 1)$
\cite{OSVPRE,OSVJSP,Hanke}. However, the results on localization and
delocalization of flat knots apply to a model which is a too
crude representation of knots in three dimensions, which are the
challenge here.

In a recent Letter~\cite{Marcone} we reported a preliminary investigation
of the size of knots in a flexible polymer ring fluctuating in
equilibrium in 3D. By modelling the ring
configurations as self-avoiding polygons (SAPs)
on a cubic lattice, we could take fully into account excluded volume.
Thanks to the inclusion of short range attractive interactions, upon lowering
the temperature $T$, the polymer ring did undergo a collapse transition
from a coil to a globule shape at the $\theta$ point
temperature \cite{Vanderzande}.

To measure the average size of the knots in the high temperature
swollen regime, we followed two different methods \cite{Marcone}.
The first one was based on the cutting-closing strategy already
outlined above. In order to test the consistency of the results we
also adopted a completely new strategy based on the entropic
competition between two knotted loops into which a ring can be
partitioned by a slip link. If each one of the two loops contains,
e. g., a prime knot, one expects a dominance of equilibrium
configurations in which one of the loops is entropically
tightened, while the other one gains almost the whole length of
the ring. It is then tempting to identify this form of tightening
of the loop with the entropic tightening of the knot it contains.
This last tightening could also be the same as that occurring
within a singly knotted fluctuating ring.

In Ref. \cite{Marcone} we gave evidence that in the swollen
regime prime knots are weakly localized with $t \sim 0.75$. This
result was obtained with the two independent methods above, by fitting the
power law behavior of the average knot length as a function of
the total ring length. Similar methods were subsequently applied 
in Ref. \cite{Virnau2005}
to an off-lattice model of open polyethylene chain, obtaining
results for the localization of the trefoil knot in qualitative
agreement with ours.

One of the aims of the present work is to address the issue of the
consistency of the method of knot localization study based on
cutting and closing and that based on entropic competition of
loops more systematically, by testing other knot types and, most
important, by analyzing more globally the probability distribution
functions (PDFs) of the knot length measured in the different
cases.

Another purpose of the present work is that of investigating the
issue of scaling corrections for ring polymers with  fixed
topology. Recently, an attempt was made to infer the localization
properties of prime knots from the scaling correction detected in
the force-extension plots of knotted polymers whose extremes are
subjected to a force \cite{Farago}. However, the results appeared
consistent with a power law behavior of the average knot length
rather different from those we detected in ref \cite{Marcone} by
our direct measurement. Moreover, the correction estimated there
appeared weaker than the correction predicted by the field
theoretical renormalization group methods for a polymer with
unrestricted topology.

A further result, first established in \cite{Marcone} is that
below the theta temperature, in the collapsed phase, knots are
delocalized $t \simeq 1$. This important conclusion, further
supported by results in Ref.
\cite{Virnau2005}, calls for a more systematic discussion in view
of the relevance of topological entanglement in globular polymers
and biopolymers.

The plan of this paper is as follows. In the next section we
describe the model and some details of the methods of simulation.
The direct measure method will be introduced in Section III and
Section IV, where we will discuss also how one can discuss the PDF
of the knot length. Section V is dedicated to the knot length
evaluation based on the entropic competition between two loops.
This second method is the only one safely applicable in the low
temperature collapsed  regime, and we will present the result
obtained for this regime in a further Section (VI). In Section
VII, we discuss the implications that the weak localization of
knots in good solvent has on the scaling behavior of the mean 
square radius of
gyration of polymer rings and on its corrections. We close in
Section VII with a general discussion of the results.

\section{MODEL AND SIMULATION METHODS}
We model  flexible ring polymers in good solvent by self-avoiding polygons
(SAPs) on the cubic lattice, i.e. closed lattice walks whose steps
can visit each edge and each vertex of the lattice at most once
\cite{Vanderzande}. Whenever necessary  we
introduce in our model an attractive interaction potential which
lowers the total energy by $\epsilon>0$ whenever two
nearest-neighbor lattice sites are visited by non-consecutive
vertices of the SAP. This attractive interaction is sufficient to
induce a theta collapse at sufficiently low temperature $T$
\cite{Vanderzande}. The Hamiltonian of a  configuration $\omega$
with $N_i(\omega)$ nearest neighbor interactions is then
$H(\omega)=-\epsilon N_i(\omega)$.
\begin{figure}[tbp]
\begin{center}
\begin{tabular}{c}
\includegraphics[scale=0.40]{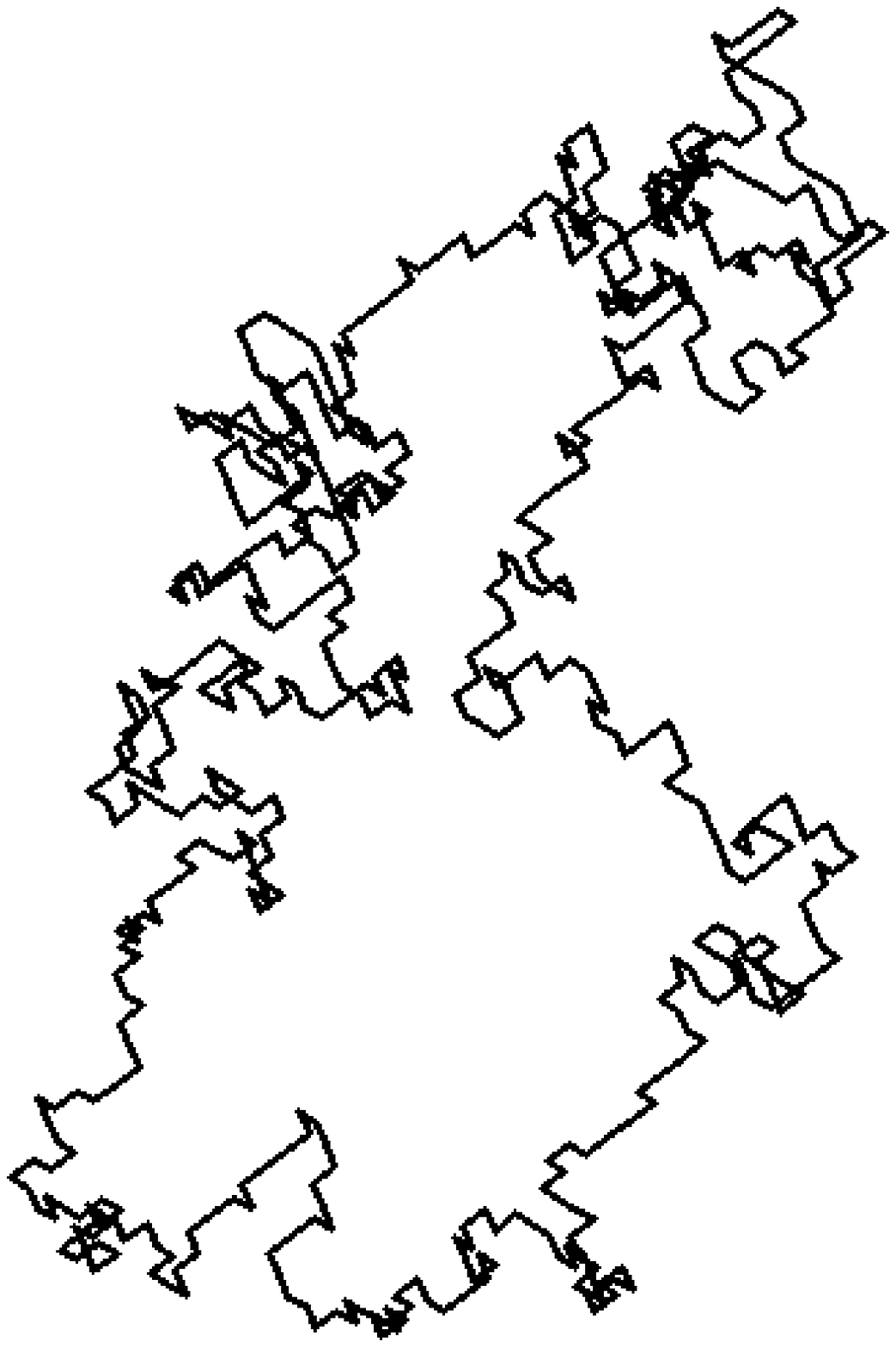}\\
\includegraphics[scale=0.40]{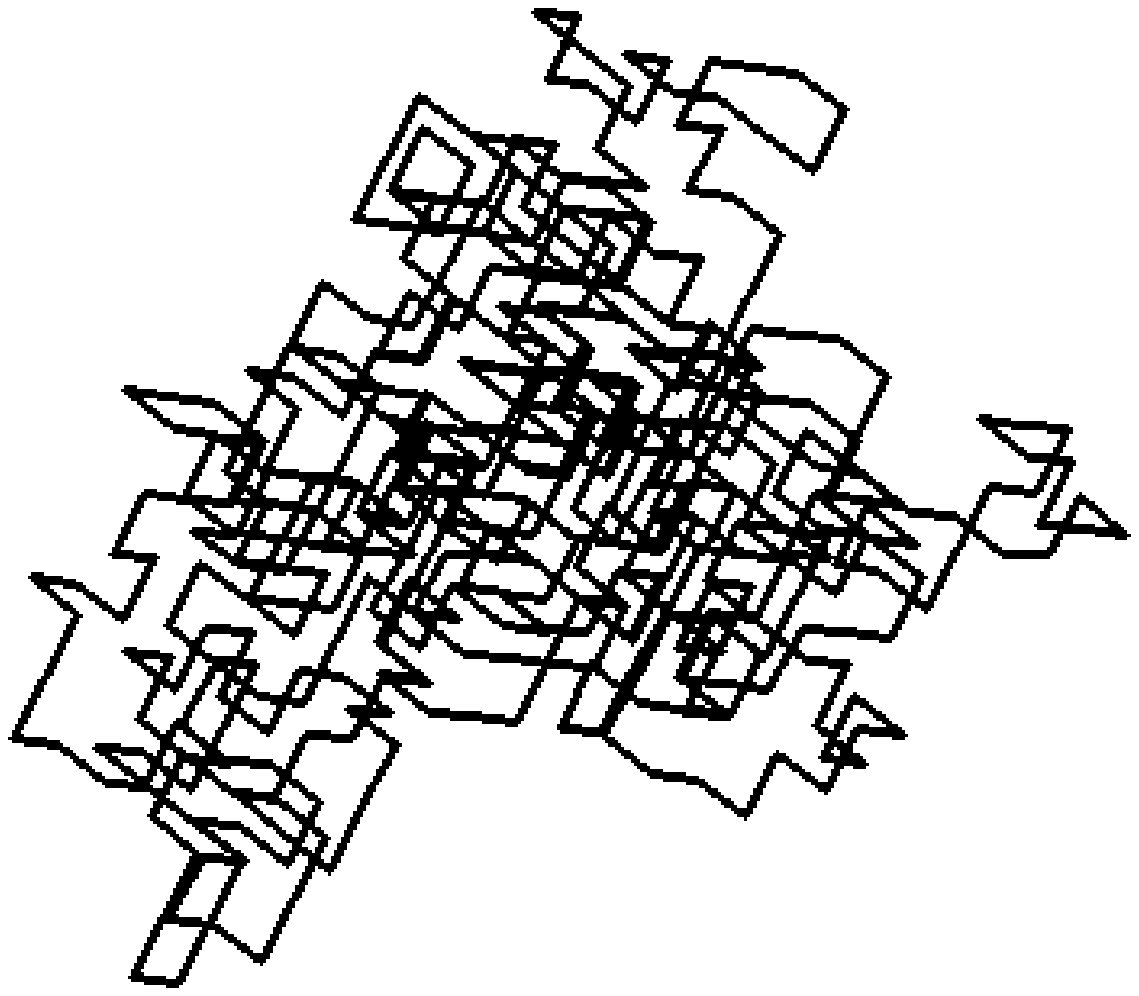}
\end{tabular}
\end{center}
\caption{Two knotted ($3_1$) SAP configurations at equilibrium
sampled by a BFACF algorithm: the top configuration refers to the
swollen regime while the bottom one to the collapsed phase.}
\end{figure}
For a fixed temperature $T$, configurations $\omega(\tau)$, with a
fixed knot type $\tau$, are sampled at equilibrium by using
a Monte Carlo approach based on the BFACF algorithm, \cite{CPS}
\cite{Sokal}. This is an algorithm which samples along a Markov
chain in the configuration space of polygons of variable $N$ and with
fixed knot type. The statistical ensemble considered is thus grand
canonical, with a fugacity $K$ assigned to each polygon step. We
adopt this algorithm because it preserves the topology and is irreducible within
each set of configuration having the same knot type \cite{Janse91}.
At a given
$T$ we used a multiple Markov chain (MMC) procedure \cite{Tesi}
in which configurations are exchanged among
ensembles having different step fugacities~\cite{Orlandini98}. This is
done in order to improve the efficiency of the sampling, especially
at low $T$. In our simulations in the swollen regime
the MMC's combined up to 10 processes at
different $K$'s ranging from $K=0.2109$ up to
$K=0.2130$~\cite{note1}. 

A relative disadvantage of the BFACF sampling method is that
correlation times are relatively long, making it difficult to
collect a sufficient statistics of uncorrelated data at large $N$.
To improve the statistics  for high values of $N$ in our
simulations of knotted SAPs, in the direct measure for the $3_1$ and
the $4_1$ knots we also made use of a different sampling procedure based on
the two-points pivot algorithm \cite{Madras}. The  two-points
pivot algorithm is known to be ergodic in the set of all SAP's
with fixed $N$ and quite efficient in sampling uncorrelated
configurations \cite{Madras}. Unfortunately, pivot moves can change
the knot type  and a check of the topology of each
sampled SAP configuration is needed. This is done  by calculating the
Alexander polynomial $\Delta(z)$ in $z=-1$ and $z=-2$
\cite{Rolfsen}. The sampled configurations are then partitioned
according to their knot type and the cut and join procedure is
performed as before. Note that, since it explores the whole space
of $N$-step SAPs, the Pivot algorithm  is not very efficient in
sampling small $N$ configurations with fixed knot type. For
example for $N=1000$ the probability of forming a knot is $\sim
0.004$~\cite{Yao01,Kantor03}  meaning that one should wait on
average $1000$ uncorrelated unknotted configurations before seeing
a knotted one. However for $N>1000$ simple prime knots start to
appear with a sufficient frequency and a reasonable statistics at
fixed knot type starts to be possible.

\section{DIRECT MEASURE OF THE KNOT LENGTH: THE CUT and JOIN PROCEDURE}
\begin{figure}[tbp]
\includegraphics[angle=0,width=\WIDTHC]{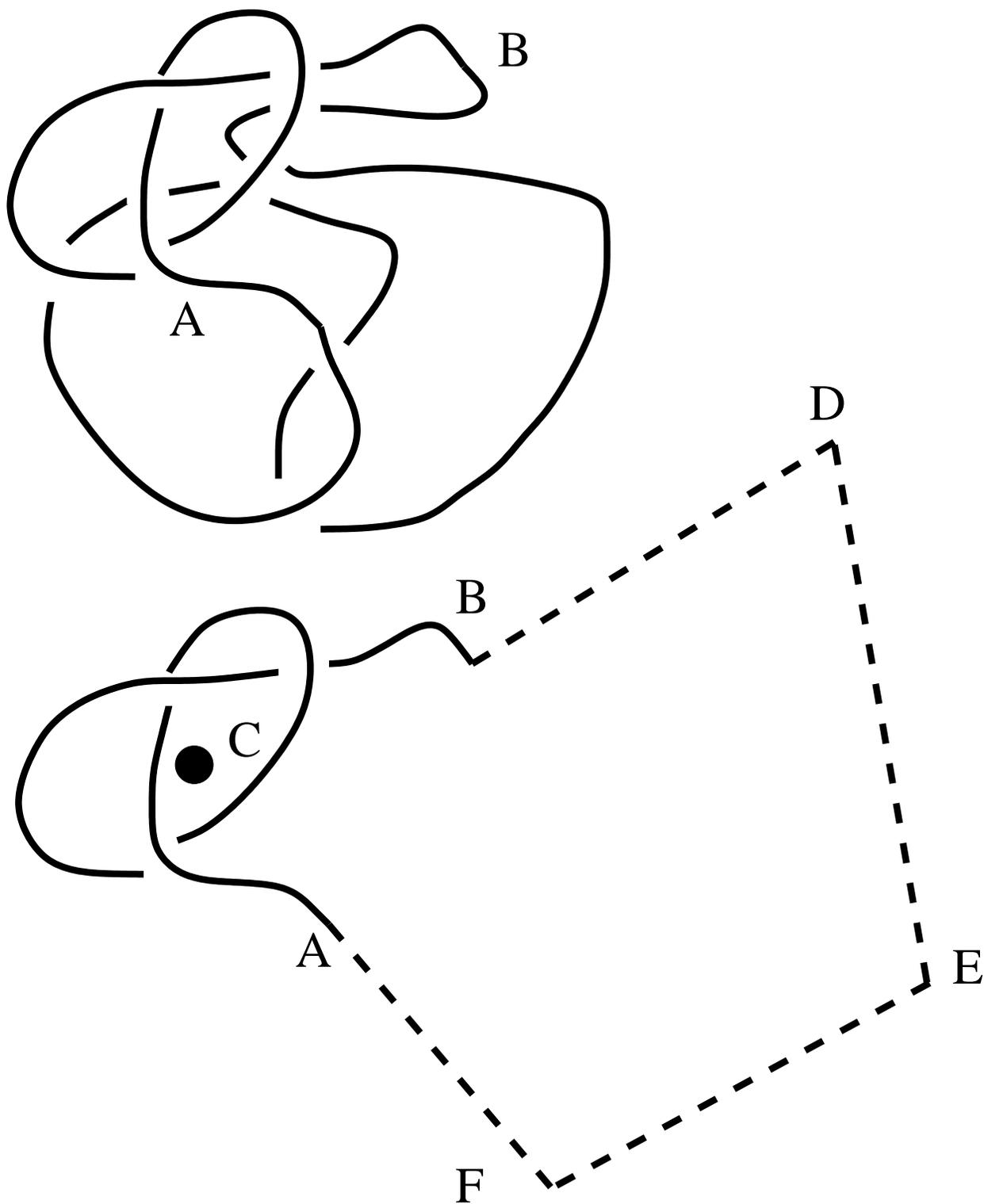}
\caption{A sketch of how the closure scheme works: 
for a given extracted arc with extremes $A$ and $B$
we compute the center of mass $C$ and construct two segments that
go far from it (they are constructed on the line connecting $C$
with $A$ and $B$). We then complete the loop by connecting the
extremes of these segments ($D$ and $F$) to a point distant from
the arc ($E$).}
\label{closure}
\end{figure}

 For each sampled polygon with fixed prime knot type $\tau$ we
measure the length $\ell$  by determining the shortest possible arc
that contains the knot. The idea is
rather intuitive and has been considered in previous works on
topological entanglements by several authors \cite{Janse92,Katrich,Millet05,Marcone}.
The ways in which this method can be
implemented can be different and may reveal very important in
order to lower systematic errors as we will discuss below. Our procedure
works as follow \cite{Marcone}: given a knotted configuration
we extract open arcs of different length
by following a recursive procedure. Each
arc is then converted into a loop by joining its ends at infinity
with a suitable path (Figure \ref{closure}) and the presence of
the original knot is checked by computing, on the resulting loop,
the Alexander polynomial $\Delta (z)$ in
$z=-1$ and $z=-2$~\cite{Rolfsen,Volog}.
Clearly, the additional
path (dotted in Fig.~\ref{closure}) can topologically interfere
with the original arc (despite the procedure
tries to avoid this as much as possible) and this could be
a source of systematic errors
(see Fig. 2). This is a disavantage common to
all the procedures that define a knotted arc by closing it
into a loop~\cite{Janse92,Katrich}. Our
goal here is to find an optimal closure procedure that minimizes
such error. Moreover, as proposed in \cite{Marcone}, we expect that
the systematic inconsistencies of which this method suffers,
should not affect the asymptotic
statistical characterization of localized knots.

A rough indication of the systematic error can be given by counting how many
times the cut and join procedure finds a knot in arcs extracted from
unknotted rings. For our algorithm this test gives a percentage of errors,
of the order of $0.2\%$,with $N=500$, quite low
if compared, for example, to the error estimate in \cite{Katrich}  for a
similar system. A possible explanation is that in \cite{Katrich} the
closure is chosen randomly, while in our case it is deterministic and
conceived to avoid the pre-existing skein as much as possible.

The procedure we follow in order to identify the shortest arc 
containing the knot for each SAP configuration, is of iterative type
and realizes a progressive reduction of the length of several 
initial trial arcs.

\section{DATA ANALYSIS AND RESULTS}
We first focus on the size of prime knots for rings at
equilibrium in the high temperature regime. In Ref. \cite{Marcone}
we gave preliminary results indicating that in 
the swollen phase the trefoil
and the figure eight knot are weakly localized.
Our aim is to
make these results more robust by considering
other prime knots. Here and in the following, brackets 
indicate fixed N averages,
obtained, whenever necessary, by a suitable binning of the data.
For a fixed knot type we sampled roughly $10^6$ uncorrelated
SAP configurations and for each one of those we estimated the size of
the hosted knot by the procedure previously described.
\begin{figure}[tbp]
\includegraphics[angle=0,width=\WIDTHC]{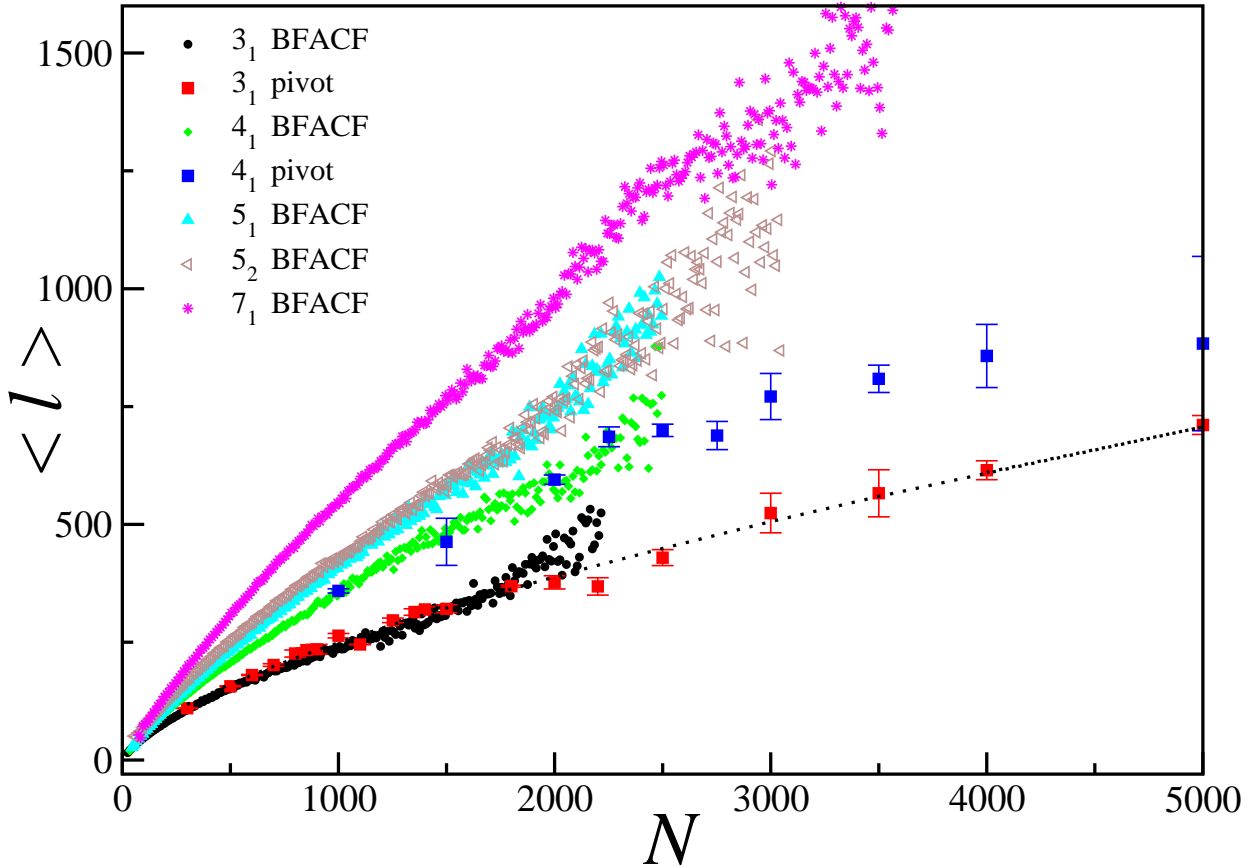}
\caption{Average knot size $\langle \ell
\rangle $ as a function of the total length of the polygon $N$ for
different prime knots. Filled symbols correspond to BFACF data,
whereas empty symbols represent determinations obtained by the pivot
algorithm. The BFACF data has been binned in $N$. The dotted curve
corresponds to a fit of the form $\langle \ell \rangle \simeq AN^{t}$ 
for the $3_1$
with $t$ given by the estimated value in Table I. Note that the
data for $5_1$ and $5_2$ are almost superimposed.} \label{diretta}
\end{figure}
In Fig. \ref{diretta} we show the $N$ dependence of the average
knot size $\langle \ell \rangle$ obtained in this way for
different prime knots. The plots give evidence that
 $\langle \ell \rangle \sim N^t$ in all cases. By performing log-log fits of the
 data and a finite size scaling analysis,
we obtain the $t$ estimates reported in table \ref{table1}.
\begin{table}
\begin{center}
\begin{tabular}{|c|c|}
\hline
\bf{prime knot} & \bf{$t$}\\
\hline
$3_1$  & $0.67\pm0.05 $  \\
$4_1$  & $0.77\pm0.07 $  \\
$5_1$  & $0.80\pm0.07$   \\
$5_2$  & $0.80\pm0.05$   \\
$7_1$  & $0.85\pm0.08$ \\
\hline
\end{tabular}
\end{center}
\caption{Estimates of the knot size exponent $t$ by
 the cut and join approach. They have been obtained by a
linear fit of the log-log plots of Fig.~\ref{diretta}. For the $3_1$ and the $4_1$ knots the estimates are based on both BFACF and pivot data.}
 \label{table1}
\end{table}
These estimates give a good evidence that prime knots are weakly
localized in the swollen regime, i.e. $t<1$. Moreover the overlaps of the plots for
$5_1$ and $5_2$.
seem to suggest that knots with the same minimal crossing number
have very close average size, even for relatively small $N$.

The possible dependence of the exponent $t$ on the knot type is
however less clear and to clarify the issue both a better sampling
at  high $N$ and a systematic analysis of finite size corrections
are needed. From Fig. \ref{diretta} one can indeed notice that for
large $N$ the BFCAF sampling technique starts to deteriorate since
the statistics becomes quite poor. This is due to the difficulty
of sampling properly the high $N$ region of the configurational
space, since the BFACF algorithm has the disadvantage of very long
autocorrelation times \cite{CPS,Sokal,Madras}. For the knots $3_1$ and $4_1$
we complemented our
BFACF determinations of $\langle \ell \rangle$ with data obtained
from the two-points pivot algorithm mentioned in Section 2. The
average knot lengths obtained with this simulation method overlap
the BFCAF estimates in Fig. \ref{diretta} for $N \leq 2500$ . The
pivot data extend up to $N \sim 3500$ and are more consistent with
the expected power law behavior in the high $N$ region. 
One can see a tendency of the exponent to grow
with increasing number of the minimal crossing number of the
knots. However, the statistical uncertainty is relatively large
and it is legitimate to suspect finite size corrections to be
stronger for more complex knots.

\begin{figure}[tbp]
\includegraphics[angle=0,width=\WIDTHC]{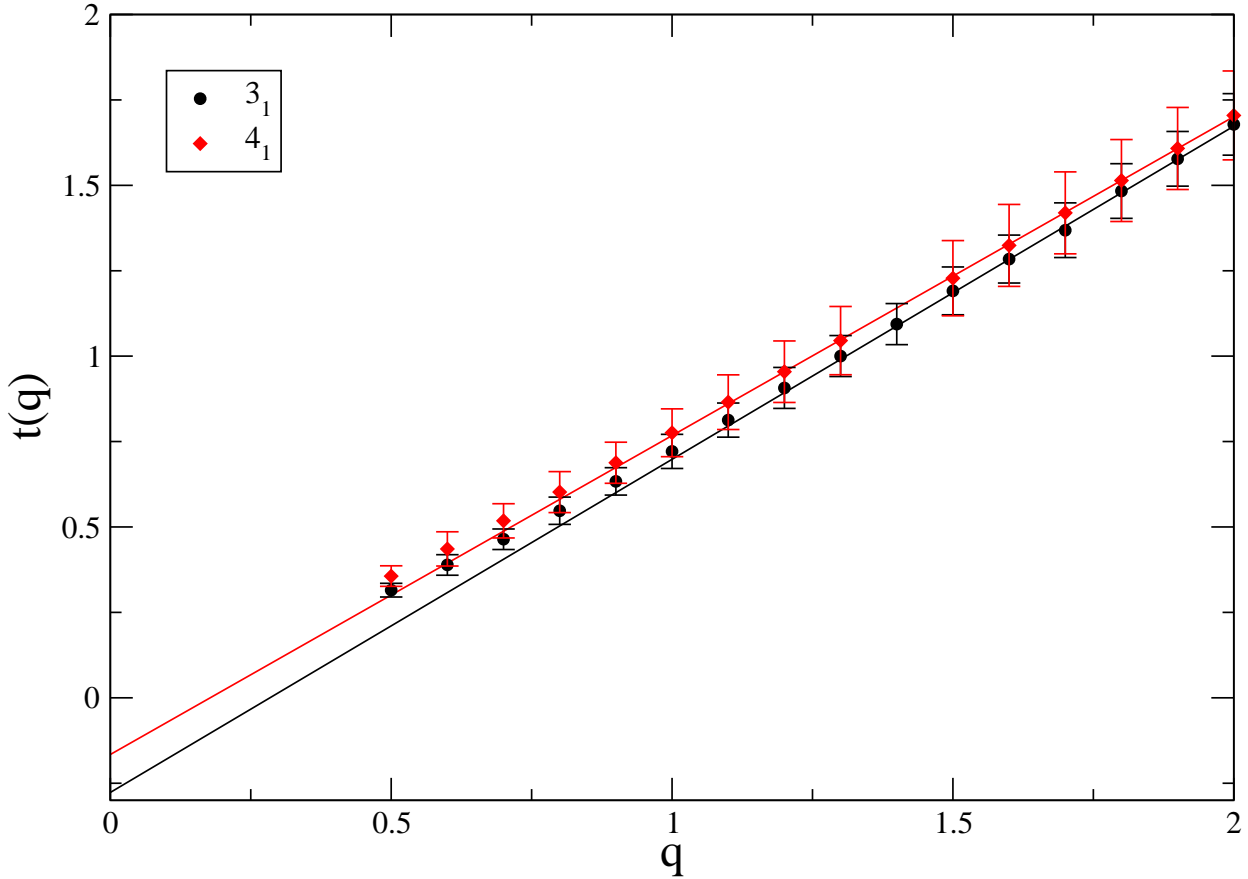}
\caption{Values of the exponents $t(q)$  as a function
of the order $q$ of the moment for the knots $3_1$ and $4_1$.
The lines are best fits for 
$1.3<q<2$. There are
deviations from the linear dependence for smaller values of $q$
due to finite size effects.}
\label{moments}
\end{figure}

As far as the scaling analysis is concerned,
a more solid and detailed control should be
achieved by analyzing the full probability distribution function
(PDF) of $\ell$ as a function of $N$ i.e. $p(\ell,N)$. In analogy
with previous works on similar problems \cite{Zhandi} \cite{Carlon}
one can assume, for the PDF, the following scaling form:
\begin{equation}
\label{pdielle} p(\ell,N)=N^{-c}f \Big(\frac{\ell}{N^D}\Big).
\end{equation}
where the scaling function $f$ is expected to approach rapidly zero as
soon as $\ell > N^D$, ($D \leq 1$). The quantity $N^D$ is a sort of cutoff on
the maximum value $\ell$ can assume. We expect $D=1$, because
there are no reason \emph{a priori} to think that there exists some
`topological cutoff' which limits the size of the knot.
Unfortunately, to look at
the scaling behavior of the PDF directly, e.g. by means of collapse plots,
 is a quite difficult task
that needs a huge amount of data and is not feasible in this
context. We can instead perform an analysis based on the scaling
behavior of the moments of the PDF in Eq. $(1)$ ~\cite{Tebaldi}.
This method relies on the following consideration: given
the scaling behavior (\ref{pdielle}) for the PDF, its $q$-th moment $(q>0)$
should obey the asymptotic law:
\begin{equation}
\langle \ell ^ q \rangle \sim N^{Dq+D(1-c)} \equiv N^{t(q)}
\end{equation}
and the two parameters $D$ and $c$ can be deduced by fitting the
the estimated exponents $t(q)$ against the order $q$ \cite{note2}
In Fig.~\ref{moments} the estimated
values of the exponent  $t$ are shown as a function of $q$ for the
prime knots $3_1$ and $4_1$. As usual in this kind of analysis
\cite{Tebaldi}, the plots of $t(q)$ show deviations from linearity
at relatively low $q$, due to finite $N$ scaling correction
effects. However, an optimal window of linearity can in general be
identified for values of $q$ which are somewhat larger, but not so large to
cause problems with the sampling of the corresponding moments due
to poor statistics. It makes sense then to rely to extrapolations
of the linear behavior within these windows for a determination
of both $D$ and $c$. Indeed, from the slope and the intercept of
these straight fitting lines we obtain the estimates given in
table~\ref{table2} for a number of different prime knots.

For $3_1$ and $4_1$ the estimates have been obtained by adding to
the BFACF data those obtained with the pivot algorithm. We notice
that $D$ is reasonably close to the value we expected ($D=1$)
especially for the $3_1$ case. The discrepancy between the
expected value and the measured one gets larger as the difficulty
of sampling at large enough $N$ increases. This sampling gets
poorer with increasing knot complexity. Indeed the most reliable
estimate of $D$ is that obtained for the trefoil knot, for which
the sampling is the best. Assuming for this knot $D=1$ and $c
=1.25$ we would obtain $t=0.75$. The estimates are all consistent
with the expectation that prime knots are weakly localized in
polymer rings in the swollen phase \cite{Marcone}. Moreover,
compared with the estimates of Table \ref{table1}, those of table
\ref{table2} vary considerably less with knot type. The results
suggest that the knot length growth exponent $t=t(1)$ for prime
knots could be independent on the knot type.

\begin{table}
\begin{center}
\begin{tabular}{|c|c|c|c|}
\hline
\bf{knot type} & \bf{D}  & \bf{c} & \bf{t}\\
\hline
$3_1$  & $0.958\pm 0.004$  & $1.25 \pm 0.04$& $0.72 \pm 0.03$\\
$4_1$  & $0.934\pm 0.004$  & $1.18 \pm 0.04$& $0.77 \pm 0.04$\\
$5_1$  & $0.918\pm 0.002$  & $1.14 \pm 0.03$& $0.79 \pm 0.03$\\
$5_2$  & $0.863\pm 0.003$  & $1.11 \pm 0.04$& $0.76 \pm 0.04$\\
$7_1$  & $0.864\pm 0.004$  & $1.06 \pm 0.09$& $0.81 \pm 0.09$\\
\hline
\end{tabular}
\caption{Results from the analysis of moments
 of the knot size PDF for the cut and join approach.}

\label{table2}
\end{center}
\end{table}

\section{KNOT LENGTH ESTIMATES BY ENTROPIC COMPETITION}
As remarked in the previous section, measures of the knot length
based on the cut and join procedure lead to systematic errors that
are somehow uncontrolled. These errors are mainly due to the
topological interference between the chosen arc and the polygonal
used to join the ends of the arc at infinity (Fig.\ref{closure}). 
This problem becomes
much more serious in the case of collapsed ring polymers since the
chance to find the ends of the arc deep inside the globule formed
by the arc itself is very high. To overcome this problem, a
completely different procedure has been recently introduced in
Ref.~\cite{Marcone}. The idea consists in partitioning a SAP into
two (mutually avoiding) loops by a narrow slip-link that does not
allow a complete migration of one loop, or of its knot, into the
other loop. The whole topology of such structure can be
characterized by the knot types $\tau_1$ and $\tau_2$,
respectively of the first and the second loop, and by the linking
state between the two loops. On this model a Monte Carlo dynamics,
based again on the BFACF algorithm, is then implemented in such a
way that the overall topology of the configuration is conserved.
In our simulations we considered only cases in which the two loops
are unlinked. Let us start considering the most symmetric
situation: $\tau_1=\tau_2$. At equilibrium, since the number of
configurations for the whole SAP is maximum when one of the loops
is much longer than the other one, most configurations break the
symmetry between the two loops showing a marked length unbalance.
Typically, in one of the two loops the knot has a very large share
of the whole SAP at its disposal, while the other loop is just
long enough to host its knot. This effect is very pronounced if
both loops are unknotted ($\tau_1=\tau_2=\emptyset$). In this case
the smaller loop is practically always confined to the minimal
length allowed by the model (Fig.~\ref{two_un}). A similar
behavior has been also found   \cite{Zhandi} for a model of
independent loops, i.e. loops for which the mutual avoidance is
neglected.
\begin{figure}[tbp]
\includegraphics[angle=0,width=\WIDTHC]{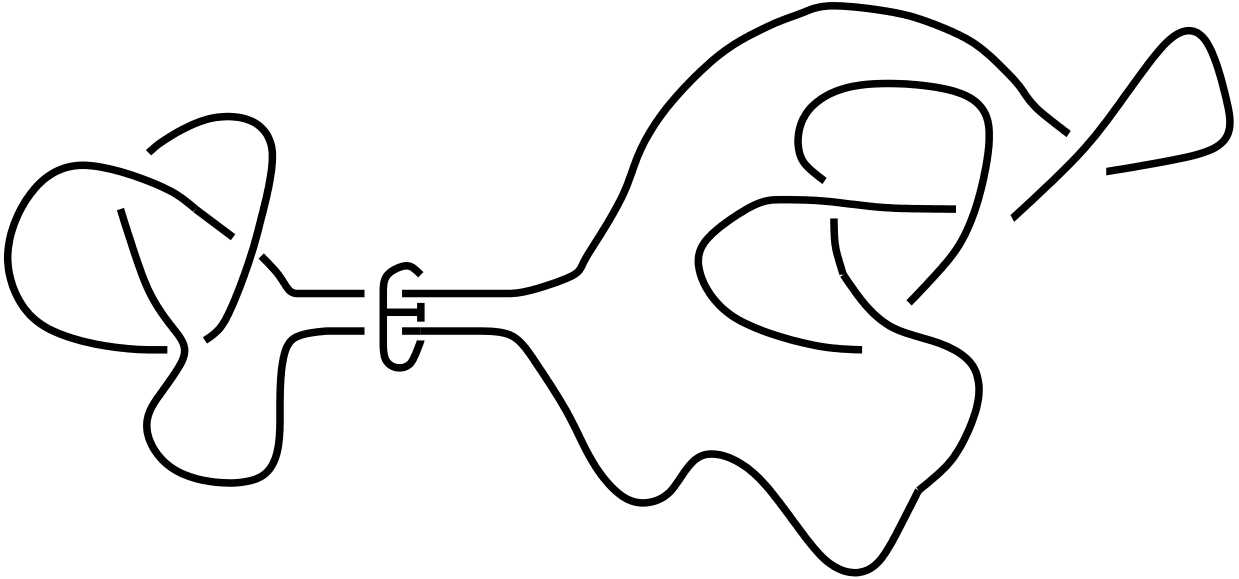}
\caption{Schetch of a trefoil knot forced to its typical length by
the competition of another knot}
\end{figure}
\begin{figure}[tbp]
\includegraphics[angle=0,width=\WIDTHC]{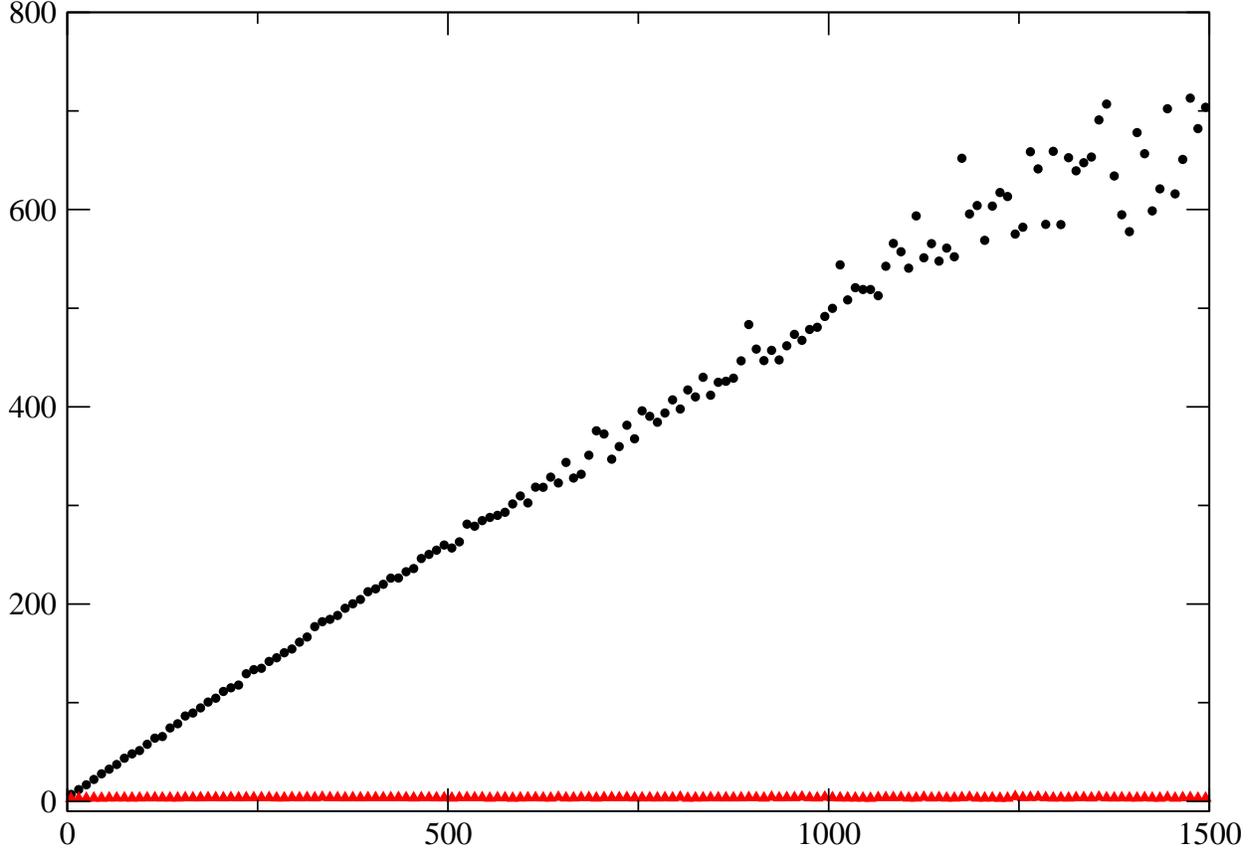}
\caption{Competition between unknotted loops: the circles represent the
length of the longer loop while the triangles the length of the shorter loop. 
While the former delocalizes, the latter seems to be
forced to its minimal length ($4$ edges).}
\label{two_un}
\end{figure}

\begin{figure}[tbp]
\includegraphics[angle=0,width=\WIDTHC]{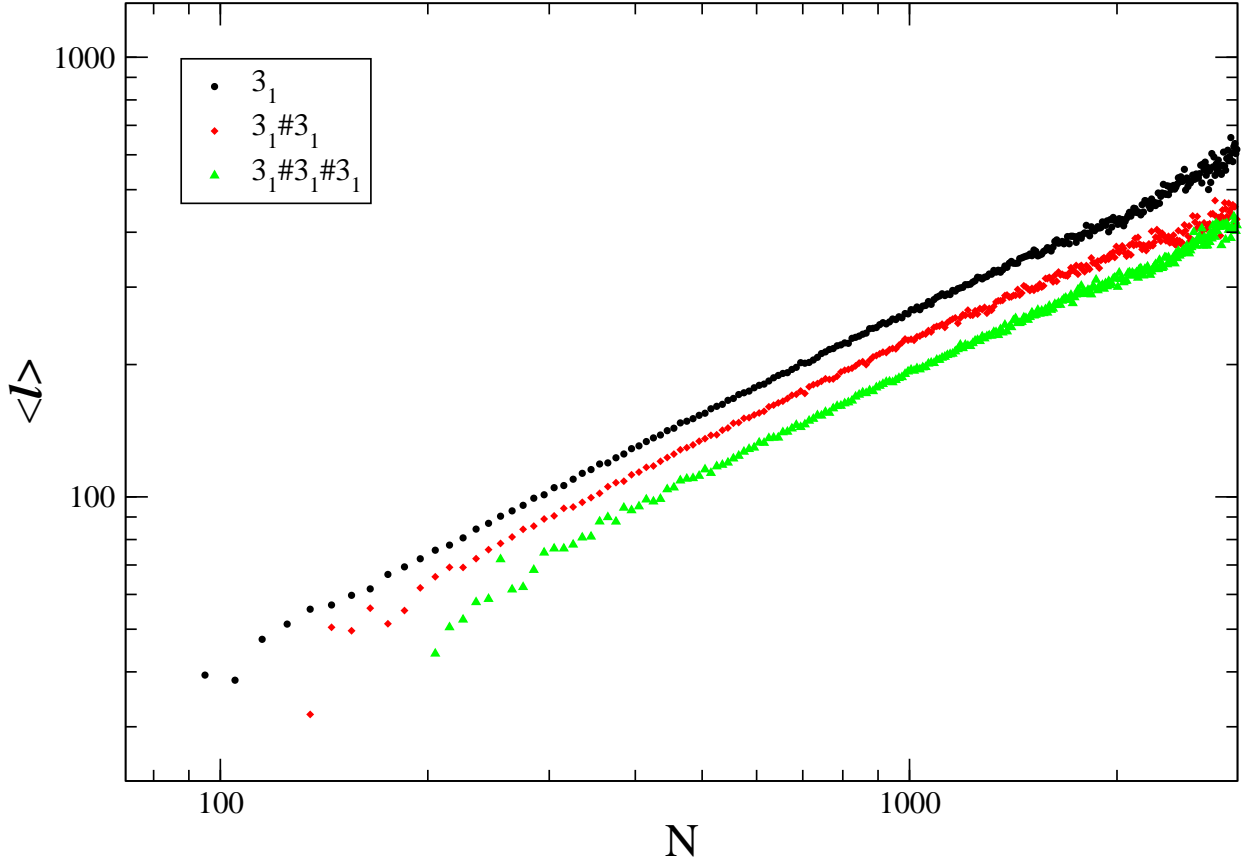}
\caption{Log-log plot of  $\langle \ell \rangle$ for the
shorter loop as a function of $N$. Different symbols
correspond to different knots $\tau$ hosted by the second loop.}
\label{allknots}
\end{figure}

Consider now the situation in which each loop hosts the same prime
knot ($\tau_1=\tau_2\equiv\tau\ne \varnothing$). Here we identify
the knot size $\ell$ with the length of the smaller loop. With a
little abuse of notation we are not going to distinguish between
the length of the knot and the length of the shorter loop, as we
will eventually argue that their scaling behaviors are the same.
So, we will refer to both of them with the symbol $\ell$.
Figure~\ref{allknots} shows the mean value of $\ell$ as a function
of the total size $N$ for the cases $\tau=3_1,4_1,5_1$ and $7_1$.
Unlike in the unknotted case, the size $\ell$ is not fixed to the
minimum value allowed by the knot type considered (for example
$23$ for $3_1$~\cite{Diao}), but fluctuates and grows according to
\begin{equation}
\langle \ell \rangle \sim N^t.
\end{equation}
A log-log fit of the data gives for $t$ estimates that are in
agreement with those obtained by the direct measure of the knot
length $\ell$ based on the cut and join method. This corroborates
the preliminary results in Ref. \cite{Marcone}. The  analysis also
confirms that the entropic competition approach is a valuable,
alternative, tool for estimating the scaling behavior of the knot
size. Note that, unlike the cut and join approach, the entropic
competition method allows to estimate also the average size of
composite knots when they are tightened close to each other within
a tight loop. In Fig.~\ref{allknots}, for example, we report the
result for the case $(3_1\#3_1,3_1\#3_1)$. It is interesting to
notice that for this and other composite knots the $N$ dependence
of $\langle \ell \rangle$ is similar to that observed for the
prime knots considered. Thus, when the components of a composite
knot are maximally localized, the exponent $t$ does not seem to
differ much from that valid for prime knot localization.

\begin{figure}[tbp]
\includegraphics[angle=0,width=\WIDTHC]{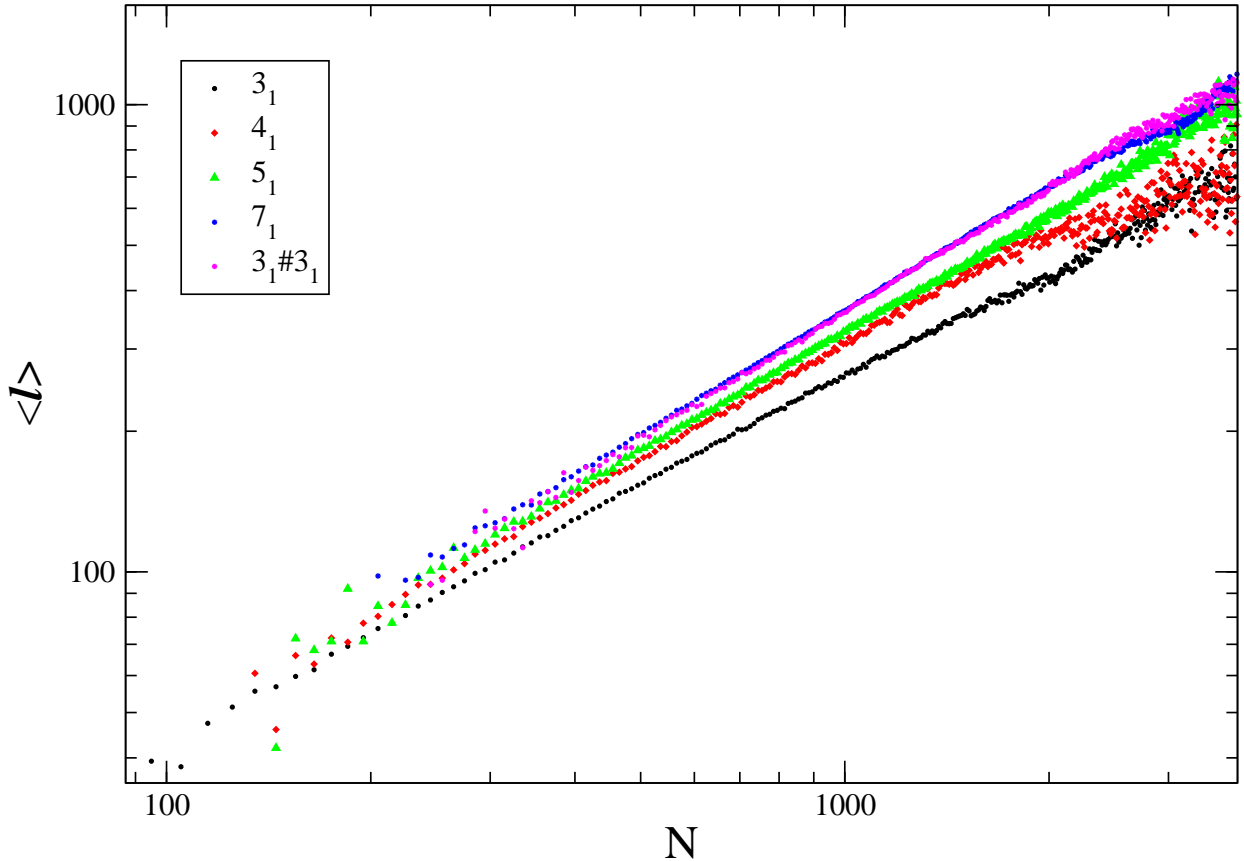}
\caption{Log-log plot of the mean length $\langle \ell \rangle$ of the
shortest loop (hosting $3_1$) as a function of $N$. Different symbols
correspond to different topologies of the longer loop.}
\label{marconefig4}
\end{figure}

Why does the entropic competition method work so well? We learned
from the cut and join approach that a knot hosted in a loop is
weakly localized, i.e. its length grows as a power law of $N$ with
exponent $t<1$. This means that the loop configuration in which the
knot has strictly its minimal length (independent on $N$) is not
the only one favoured entropically. In a system of two equally
knotted loops one of the loops will always grow as $N$ for the
same entropic reasons we discussed in the case of two unknotted
loops. Now, however, the smaller loop is knotted and since the
knot tends to be weakly localized, it forces the whole loop to
behave in the same way. Since the two loops host the same knot
type, the situation is perfectly symmetric and the system chooses
spontaneously which of the two loops to make longer. If, on the
other hand, we break explicitly the loop symmetry by inserting
different knots in the two loops, ($\tau_1\ne\tau_2$), the system
at equilibrium tends to have as the smaller loop the one hosting
the simpler of the two knots. The entropic argument described
above for the smaller loop is still valid here and we do not
expect changes in the scaling behavior of $\langle \ell\rangle$.
This is indeed the case as one can see from Fig.
\ref{marconefig4}, which shows the $N$ dependence of $\langle \ell
\rangle$ in the cases in which the simplest knot is the trefoil
($\tau_1 = 3_1$). In this case the average of $\ell$ reported has
been based on sampling the length of the loop with knot $3_1$ only
in configuration in which the same loop is the smaller one. One
can notice that, as the complexity of $\tau_2$ increases, the
minimal length $\ell_{min}$ to host such knot increases and, for
fixed $N$, there would be less edges at disposal of the knot
$3_1$. In other words the increase of the knot complexity in the
longer loop corresponds to an increase of the ``entropic force''
it applies on the smaller loop. This action, however, affects only
the amplitude of the scaling behavior of $\langle \ell \rangle$,
keeping the exponent $t$ unaltered as one can guess from the
slopes of the log-log plots in Fig. \ref{allknots}.

As in the case of the direct measure, robust estimates of $t$ can
be obtained by performing the analysis of the moments if the
probability distribution function $p(\ell,N)$, where $\ell$ is the
length of the shorter loop in the case of competition between two
equal prime knots, or the length of the loop hosting the simpler
knot (when it is also the shorter) in the case of two different
competing knots. The obtained estimates for the exponents of $p$
are reported in Table \ref{table3}. As one can see, they are
compatible with those presented in Table \ref{table2}, obtained
from the cut and join method.
\begin{table}
\begin{center}
\begin{tabular}{|c|c|c|c|}
\hline
\bf{$(\tau_1,\tau_2)$} & \bf{D} & \bf{c} & \bf{t}\\
\hline
$3_1,3_1$ & $0.939 \pm 0.002$ & $1.28 \pm 0.03$ & $0.68 \pm 0.02$\\
$3_1\#3_1,3_1$ & $0.916 \pm 0.002$ & $ 1.30 \pm 0.03$ & $0.64 \pm 0.02$\\
$3_1\#3_1\#3_1, 3_1$ & $0.940 \pm 0.005$ & $1.33 \pm 0.04$ & $0.63 \pm 0.03$\\
$4_1,4_1$ & $0.892 \pm 0.006$ & $1.20 \pm 0.07$ & $0.82 \pm 0.07$\\
$5_1,5_1$ & $0.939 \pm 0.002$ & $1.14 \pm 0.05$ & $0.81 \pm 0.04$\\
$7_1,7_1$ & $0.937 \pm 0.004$ & $1.13 \pm 0.08$ & $0.82 \pm 0.07$\\
$3_1\#3_1,3_1\#3_1$ & $0.940 \pm 0.002$ & $1.07 \pm 0.06$ & $0.87 \pm 0.06$\\
\hline
\end{tabular}
\end{center}
\caption{Estimates, with the method of moments,
of the knot size exponent $t$ obtained by looking at the average
size of the smallest loop of the two loops model.
}
\label{table3}
\end{table}

\section{KNOT SIZE IN  COLLAPSED POLYGONS}
If in the SAP model we introduce an attractive interaction between
non consecutive n.n. monomers, we mimic the effect of a bad
solvent. In this case, upon lowering the temperature below
$T_\Theta$, the SAP undergoes a collapse transition
\cite{Vanderzande} from a swollen to a collapsed phase. It is
interesting to see how the degree of localization of knots depends
on the quality of the solvent. We are interested in determining
how the size of the knot $\ell$ behaves for highly condensed
polygons. Previous studies on flat knots \cite{OSVPRE}\cite{Hanke}
showed that in the compact regime they delocalize. A similar
delocalization was first predicted in Ref. \cite{Marcone} for real
$3d$ knots. Unfortunately an estimate of $\langle \ell \rangle$
obtained by a cut and join method would not be reliable for
compact configurations since the cut and close procedure would
alter with high probability the topology of the chosen arc
~\cite{note3}.

\begin{figure}[tbp]
\includegraphics[angle=0,width=\WIDTHC]{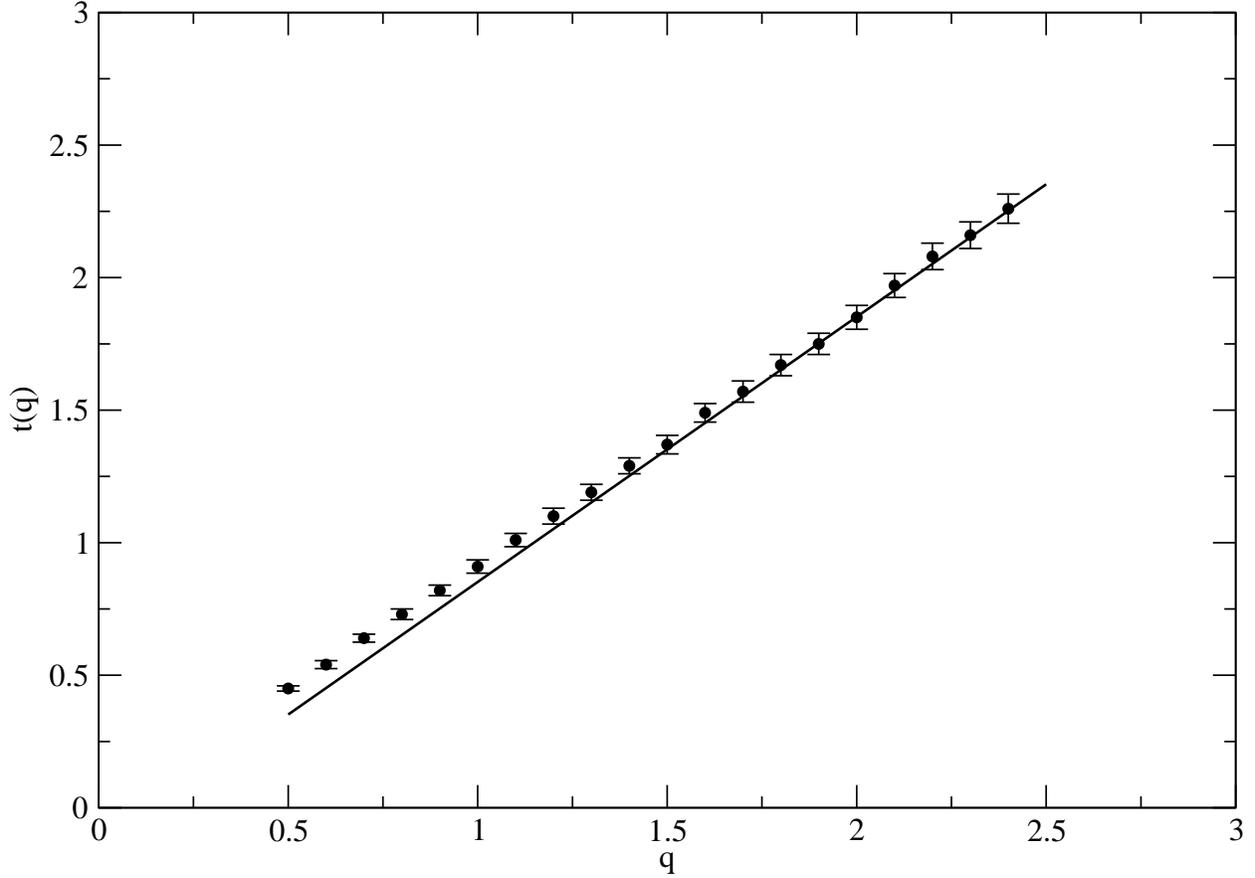}
\caption{Exponent $t(q)$ in the competition between two trefoils
in the compact phase. The data for the figure eight knots overlap
those for trefoils, and are not included.} \label{lowtemptrefoils}
\end{figure}

To the contrary, the strategy based on entropic competition
does not involve alterations of the topology and
should work also for very dense configurations.
To obtain compact configurations
we have simulated the two loop model 
at $T\sim 0.53 T_\Theta$, i.e. well inside the collapsed phase.
Unfortunately, to sample SAP's below the
$\Theta$ point is in general a difficult task to achieve \cite{Tesi}.
The situation is even more delicate for grand canonical algorithms (such as
the BFACF) since, at $T<T_{\theta}$,
as the critical edge fugacity is approached
from below, the grand-canonical average number of
SAP edges undergoes a
first order infinite jump from a finite value,
rather than growing continuously to infinity as
in the $T>T_{\Theta}$ case.
In spite of this difficulty, we have been able to sample
$\sim 10^6$ uncorrelated configurations for each topology
considered and for $N$ up to $\sim 700$. In Fig.~\ref{lowtemptrefoils}
the $N$ dependence of $\langle \ell \rangle$ is reported
for the topologies $(3_1,3_1)$ and $(4_1,4_1)$. For comparison the data
coming from the swollen regime are also reported. The
difference between the two regimes is evident and a linear behavior
for the compact case can be easily guessed. Indeed, a simple linear fit
of such data gives a good correlation coefficient ($r=0.9993$)
and a slope $A_{3_1} =0.34$.

\begin{figure}[tbp]
\includegraphics[angle=0,width=\WIDTHC]{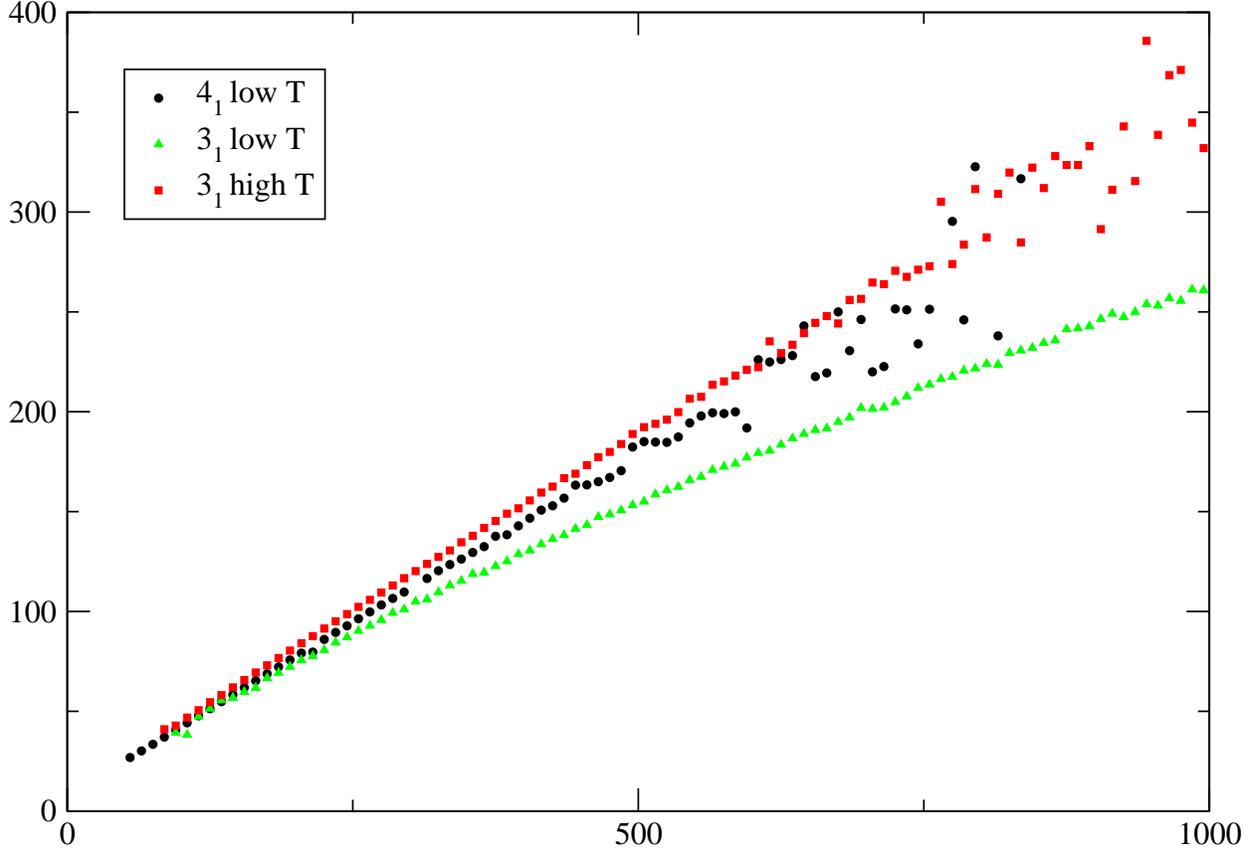}
\caption{$N$ dependence of the average size of the shortest loop,
 $\langle \ell \rangle$, for the two loops model.
The top curves corresponds respectively to the $(4_1,4_1)$ (Circles) and
and to the $(3_1,3_1)$ (Squares) topologies for $T<T_{\Theta}$.
The bottom curve has been introduced for comparison and
corresponds to the case $T>>T_{\Theta}$
for the topology $(3_1,3_1)$. }
\label{compact}
\end{figure}

An analysis of the moments of $p(\ell,N)$ confirms the conclusion
that the knots are delocalized in the globular phase. Indeed, e.g.
in the case of the $3_1$ knot, we obtain $D \simeq 0.98$ and $c
\simeq 1.1$, from plots of $t(q)$ (Fig. \ref{compact}) and
this shows that the growth of the smaller loop is linear in the
total length. A similar analysis for the figure eight case gives
$D \simeq 0.98$ and $c \simeq 1.1$, again consistent with the
expected delocalization ($t\simeq 1$).


\section{THE MEAN RADIUS OF POLYGONS WITH FIXED KNOT TYPE: CORRECTION TO SCALING.}
In this section we show how the weak localization property
of knots in the swollen regime
can have relevant consequences for the scaling behavior
of the mean squared radius of gyration of SAP's with fixed knot type.
According to the modern theory of critical phenomena based on the
renormalization group (RG), its averaged large $N$ behavior
for a ring polymer is expected to be
\begin{equation}
\langle R_g^2\rangle_N \sim A N^{2 \nu} \left ( 1 +B N^{-\Delta}
+o(N^{-\Delta})\right)
\label{r2scal}
\end{equation}
where the exponents $\nu$ and $\Delta$ are expected to be universal in the good
solvent regime. They have been estimated as $\nu = 0.5882 \pm 0.0010$ and
$\Delta = 0.478 \pm 0.010$
using field theoretic RG techniques (\cite{Guida}, see also \cite{Leg}),
consistent with the best available numerical estimates
for lattice self-avoiding walks as given by Li \emph{et al.} \cite{Li}:
$\nu = 0.5877 \pm 0.0006$, $\Delta = 0.56 \pm 0.09$.

These results are valid for phantom ring polymers with
unrestricted topology, which are the only ones that can be treated
on the basis of field theoretical RG methods.
For a ring polymer with a fixed prime knot $\tau$, it is
reasonable to expect an asymptotic form of $\langle
R_g^2\rangle_{\tau,N}^{1/2}$ like that in Eq. (4), but possibly
with different, $\tau$-dependent amplitudes. Since the $\nu$
exponent is determined by the fractal structure of the polymer
conformations, we do not expect it to change as a consequence of a
global restriction to a specific knot topology. To the contrary,
for $\Delta$ we expect the possibility of a deviation from the
value reported in Eq. (5). Indeed, for the case of prime knots, we
have established above a weak localization in the good solvent
regime. This weak localization implies the existence of a
characteristic length, $\langle \ell \rangle ^ \nu \sim N^{t \nu}
$, diverging with a power of $N$ which is subleading with respect
to $N ^ \nu$ ($t<1$). This gives the possibility of a scaling
correction exponent $\Delta' = 1-t $, as we argue below.

Let us write
\begin{equation}
\langle R_g ^2 \rangle_{\tau,N} = A_\tau N^{2\nu}\left [ 1 + B_\tau N^{-\Delta_\tau} + o(N^{-\Delta_\tau})\right] .
\label{r2scala}
\end{equation}
for the asymptotic behavior of the mean square radius of gyration
of a ring with fixed prime knot, $\tau$, in the swollen regime. In
this expression we allow for a dependence of $A$, $B$ and $\Delta$
on the type of knot. However, as far as $A$ is concerned, we
checked that the dependence on $\tau$ is very weak.

For SAP's with fixed knot type $\tau$, let $\langle \ell
\rangle_\tau$ indicate the average size of the hosted knots. In
general, if  $\langle \ell \rangle_\tau = o (N)$, as $N\to\infty$,
$\langle R_g ^2 \rangle_{ \tau ,N}$ should scale as the size of an
unknotted loop with length  $N-  \langle \ell \rangle_\tau$, i.e.
\begin{equation}
\langle R_g^2 \rangle _{\tau,N} \sim \langle R_g^2 \rangle
_{\varnothing, N-\langle \ell \rangle} \label{r2scalknot}
\end{equation}

Our estimates of $\langle \ell \rangle_\tau$ suggest
$\langle \ell \rangle_\tau \sim a_\tau N^{t}$ with $t\sim 0.75$,
roughly $\tau$-independent. 
By plugging this behavior in Eqs. (\ref{r2scala}) and (\ref{r2scalknot})
we obtain

\begin{equation}
\langle R_g ^2 \rangle _{\tau,N} = A_\varnothing N^ {2\nu} [ 1+B_ \varnothing N^{-\Delta_ \varnothing} -C_\tau N^{t-1}+ ...]
\label{r2scalc}
\end{equation}
where $C_\tau =\nu a_\tau$.

Eq. (\ref{r2scalc}) implies that either $\Delta_ \varnothing$, or
$(t-1)$ is the exponent describing the scaling correction for a
ring with fixed prime knot type $\tau$, independent of $\tau$.
Below we indicate by $\Delta ' $ this correction exponent, and it
will turn out $\Delta ' = 1-t$ or $\Delta ' = \Delta_\varnothing$,
if $1-t<\Delta _\varnothing$ or $\Delta_\varnothing <1-t$,
respectively. In any case, the fact that $1-t \sim 0.25$ tells us
that $\Delta '$ can not coincide with $\Delta \sim 0.5$ of the
phantom polymer.

Below we provide evidence that indeed $\Delta' =1-t \simeq 0.25$ is quite plausible.
At the same time one should conclude that, either $B_\varnothing = 0$, or
$\Delta _\varnothing > 1-t $.

\begin{figure}[tbp]
\centering
\includegraphics[angle=0,width=\WIDTHC]{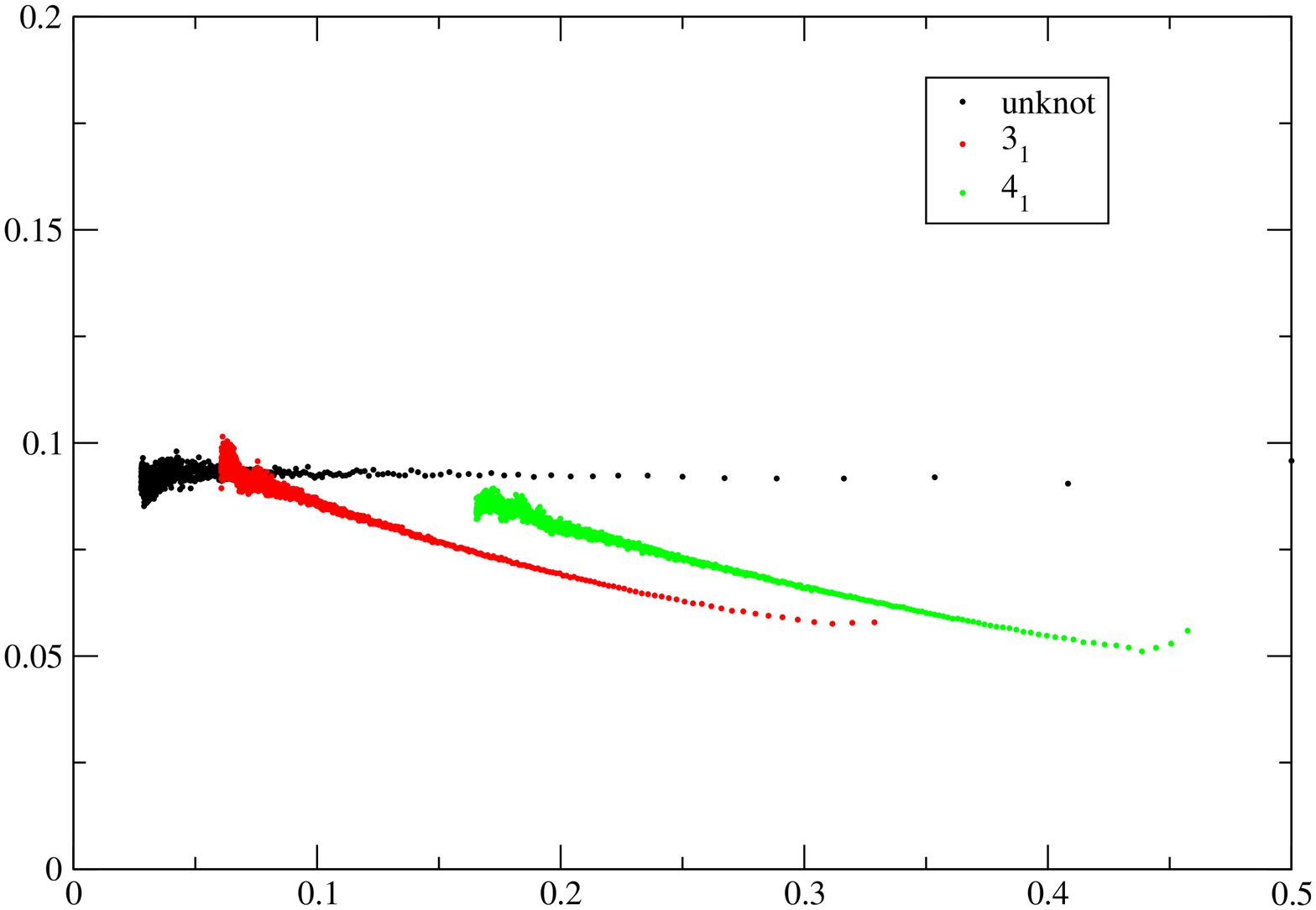}
\label{confronto}
\caption{Plot of $R^2_g/N^{2\nu}$ against $N^ {-\Delta '}$ for the unknot ($\varnothing$),
$3_1$ and $4_1$. $\Delta' = 0.25$ is considered.}
\end{figure}

In Ref. \cite{OTJW98}, the issue of the scaling of $\langle R_g^2
\rangle_\tau$ was addressed without introducing the concept that
knot localization could introduce a scaling correction exponent. A
huge collection of data was analyzed by assuming also for the
restricted knot topology the same correction exponent $\Delta \sim
0.5$ predicted for the unrestricted case. In this way a rather
convincing confirmation of the independence of $\nu$ and $A$ of
$\tau$ was obtained. To test the presence of a correction exponent
$(1-t)$ we have replotted the data of Ref.~\cite{OTJW98}, for
$\langle R_g^2 \rangle_\tau/N^{2\nu}$ assuming the scaling form
(\ref{r2scalc}) (see Figure~\ref{confronto}) with a leading
correction $ \sim N^{-\Delta '}$ and with $\nu \simeq 0588$.
Assuming a $\Delta ' \sim 1-t <0.5$, the curves appear now more
straight, as they should asymptotically, and extrapolate
 more clearly to a unique intercept with the ordinate axis,
which estimates the common, $\tau$-independent, amplitude in
Eq. (\ref{r2scalc}). The fact that for the unknot the plot is almost
horizontal suggest that either $B_\varnothing =0$, or $\Delta
_\varnothing$ is  sensibly larger than $1-t$. This becomes more
clear if one plots, on the same figure, the $N$ dependence of
$\langle R_g ^2 \rangle_{3_1,N}/N^{2\nu}$ by using different
correction terms. As one can see the data rescaled with  $\Delta
'= 1-t\sim 0.25$ are clearly more on a straight line than those
rescaled with $\Delta=0.5$ or with a much stronger, hypothetical,
correction $\Delta = 0.1$.
\begin{figure}[tbp]
\includegraphics[angle=0,width=\WIDTHC]{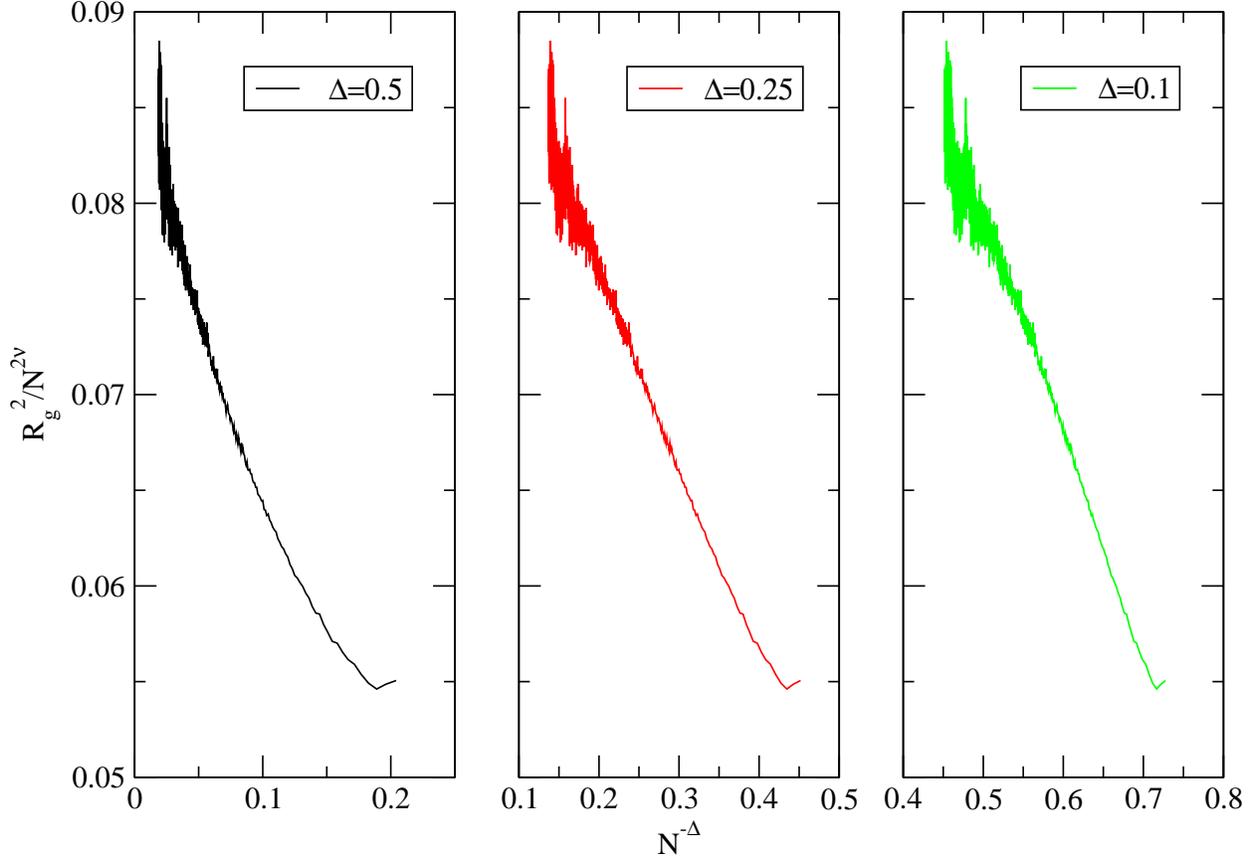}
\caption{Plot of $R^2_g/N^{2\nu}$ versus $N^{-\Delta}$ for three different
values  of $\Delta$ for $3_1$.
>From left to right: the standard value for
$\Delta$ used for ensembles  with unrestricted topology ($\Delta=0.5$),
the correction coming from our data for polymers with a prime knot in it
($\Delta=0.28$), and then a greater correction ($\Delta=0.1$). As one can
notice, the best correction seems to be given by the intermediate value
between the three proposed. These data are the ones proposed in Figure 7
 of Ref.\cite{OTJW98}.
\label{confronto2}
}
\end{figure}

In table \ref{table4} are reported the estimates of the amplitudes $A_\varnothing$
corresponding to the unknot, and to the
$3_1$ and $4_1$ knots, extrapolated from
plots like those in Fig. \ref{confronto2}, 
in the cases $\Delta ' =0.10$, $\Delta ' =0.25$,
and $\Delta ' = 0.5$. The fact that for $\Delta '=0.25$ there is an optimal agreement
among the three values further supports our conclusions on $\Delta '$.

\begin{table}
\begin{center}
\begin{tabular}{|c|c|c|c|}
\hline
$\Delta$ & \bf{unknot}  & $3_1$ & $4_1$\\
\hline
$0.1$  & $0.101\pm 0.004$  & $0.172 \pm 0.004$& $0.152 \pm 0.003$\\
$1-t$  & $0.102\pm 0.004$  & $0.110 \pm 0.004$& $0.112 \pm 0.004$\\
$0.5$  & $0.102\pm 0.004$  & $0.108 \pm 0.005$& $0.095 \pm 0.005$\\
\hline
\end{tabular}
\caption{Estimates of the amplitude $A(\varnothing)$ in eq. \ref{r2scal}
for different knot type and with $\nu(\varnothing) = 0.588$. Different
estimates correspond to different value of the correction to scaling
exponent $\Delta$. For the unknot the values $ \Delta=1-t$ and 
$\Delta=0.5$ coincide.}
 \label{table4}
\end{center}
\end{table}

\section{DISCUSSION}
In this work we addressed the problem of localization of knots in flexible
ring polymers modelled by SAPs on cubic lattice.

In the swollen regime we showed that a statistical method of prime
knot length determination based on isolating different portions of
the SAP as candidates to host the knot is consistent with an
alternative criterion, based on the entropic competition between
two knotted loops within the same ring. By a systematic analysis
of the moments of the knot length PDF of
different knots, we gave strong indication that the localization
of a prime knot is characterized by an exponent $t \sim 0.75$
describing how the average length grows as a function of $N$. The
exponent $t$ could be universal for different prime knots, or even
for composite knots whose components are tightened to remain close
to each other in the same loop.

We have shown that the weak localization of a prime knot in a
swollen ring determines a peculiar scaling correction exponent
$\Delta ' = 1-t \simeq 0.25$ for the asymptotic scaling of the
radius of gyration. This exponent implies a stronger correction
compared to that occurring for phantom ring polymers with
unrestricted knot type. We think that recent works in the
literature, addressing subtle issues concerning the scaling of
polymer rings with fixed topology \cite{Dobay}, could sharpen
their conclusions by taking into account in the numerical analysis
the different scaling correction identified here.

A remarkable advantage of the method of entropic competition between knotted loops
is the possibility of dealing with the collapsed regime without risking to suffer
too strong systematic errors in the determination of the knot length. Thanks to
a systematic analysis of data we could conclude that both prime and composite
knots fully delocalize in the globular phase, confirming a previous prediction
by the authors of the present work \cite{Marcone}.

\section{Aknoledgment} This work was supported by FIRB01 and  MIUR-PRIN05.


\begin{thebibliography}{99}
\bibitem{note2} Since for the distribution in Eq. (1) it is found in
general that $t(q)=0$ for $q \leq q_c$ with $q_c>0$, the
normalization constraint $\int p(\ell,N)=1$ does not imply
$D(1-c)=0$ i. e. $c=1$.
\end{thebibliography}
\end{document}